\documentclass[12pt]{iopart}
\usepackage{graphicx}
\usepackage{wasysym}
\begin{document}

\title{Modification of the aging dynamics of glassy polymers due to a temperature step}

\author{Mya Warren}
\ead{mya@phas.ubc.ca}
\author{J{\"o}rg Rottler}%

\address{Department of Physics and Astronomy, The University of
British Columbia, 6224 Agricultural Road, Vancouver, BC, V6T 1Z1,
Canada}

\begin{abstract}
Molecular dynamics simulations are used to investigate the connection
between thermal history and physical aging in polymer glasses, in
particular the effects of a temperature square step. Measurements of
two-time correlation functions show that a negative temperature step
causes ``rejuvenation'' of the sample: the entire spectrum of
relaxation times appears identical to a younger specimen that did not
experience a temperature step.  A positive temperature step, however,
leads to significant changes in the relaxation times. At short times,
the dynamics are accelerated (rejuvenation), whereas at long times the
dynamics are slowed (over-aging). All findings are in excellent
qualitative agreement with recent experiments. The two regimes can be
explained by the competing contributions of dynamical heterogeneities 
and faster aging dynamics at higher temperatures. As a result of this
competition, the transition between rejuvenation and over-aging
depends on the length of the square step, with shorter steps causing
more rejuvenation and longer steps causing more over-aging. Although
the spectrum of relaxation times is greatly modified by a temperature
step, the van Hove functions, which measure the distribution of
particle displacements, exhibit complete superposition at times when
the mean-squared displacements are equal.

\end{abstract}
\pacs{81.40.Lm, 81.40.Lg, 83.10.Rs}

\maketitle

\section{Introduction}

Glassy systems include such diverse materials as network glasses,
polymer glasses, spin glasses, disordered metals and many colloidal
systems. These materials present great challenges to theory and
simulation due to a lack of long-range order and very slow,
non-equilibrium dynamics.  An intriguing consequence of these ``glassy
dynamics'' is that the properties of glasses are not stationary but
depend on the wait time $t_w$ elapsed since vitrification: this
phenomenon is usually called physical aging. Polymer glasses are
particularly suited to studies of aging, since they exhibit
comparatively low glass transition temperatures, and therefore
thermally activated aging dynamics are measurable over typical
experimental timescales. One of the most comprehensive studies of
aging in polymer glasses was presented by Struik \cite{Struik}, who
showed that after a rapid quench from the melt state, there is a
particularly simple evolution of the properties of the glass. The
thermodynamic variables such as the energy and the density evolve
logarithmically with wait time, whereas the dynamical properties such
as the creep compliance exhibit scaling with the wait time
$t/t_w^\mu$, where the exponent $\mu$ is approximately unity. This
scaling law has also been confirmed for many other glassy systems
\cite{Berthier_book}, but a clear molecular level explanantion of this
behavior is still being sought.

The intrinsic aging dynamics can be modified through the influence of
stress and temperature
\cite{Struik,McKenna_JPhys15,Warren_PRE07,Rottler_PRL95,Lacks_PRL93,Lequeux_JSM,Lequeux_PRL89}. In
polymer glasses, application of a large stress usually has the effect
of ``rejuvenating'' the glass: the entire spectrum of relaxation times
is rescaled to shorter times, and closely resembles the spectrum of a
younger state of the unperturbed glass
\cite{Struik,McKenna_PES,Warren_PRE07}. The term rejuvenation
originated with Struik's hypothesis that the application of stress
increases the free-volume available for molecular rearrangements and
actually results in an erasure of aging \cite{Struik}. This hypothesis
is still somewhat controversial \cite{McKenna_JPhys15,Struik_POLY38},
however the term persists and has come to denote any acceleration of
the intrinsic aging dynamics. In colloidal systems, it has recently
been shown that stress can lead to an apparent slowing down of the
dynamics (over-aging) as well \cite{Lacks_PRL93,Lequeux_PRL89}. The
effects of temperature on the aging dynamics are similarly complex. It
is well known that glasses age faster at higher temperatures; however,
simply taking this fact into account is not sufficient to explain the
modifications of the dynamics due to changes in temperature in the
glassy state \cite{Struik}. Glasses have been shown to retain a memory
of previous annealing temperatures \cite{Bellon_EPL}, therefore, the
aging dynamics generally depends on the entire thermal history of the
sample \cite{Kovacs_APS}. Recently, a set of detailed experiments were
performed on polymer glasses that experienced a temperature square
step after a quench to the glassy state \cite{Lequeux_JSM}. Results
showed that the simple rescaling of time with $t_w^\mu$ no longer
described the creep compliance curves, and there are significant
changes to the entire spectrum of relaxation times in comparison to a
sample that did not experience a temperature step.

In a previous paper, we used molecular dynamics simulations to
investigate the dynamics of a model polymer glass under a simple
temperature quench followed by a creep experiment
\cite{Warren_PRE07}. It was shown that the model qualitatively
reproduces the effects of aging on the creep compliance, including the
phenomenon of mechanical rejuvenation at large stresses. Additionally,
through measurements of the two-time particle correlation functions,
we showed that the aging dynamics of the creep compliance exactly
corresponds to the evolution in the cage-escape time (the time-scale
for particles to undergo a significant rearrangement leading to a new
local environment). In this work, we extend our investigation of the
dynamics of aging in polymer glasses by examining the effects of a
short temperature square step, modeled after the protocol of the
experiments presented in ref.~\cite{Lequeux_JSM}. In agreement with
experiment, our simulations indicate that a downward temperature step
causes rejuvenation of the relaxation times, but an upward step yields
rejuvenation (faster dynamics) at short timescales, and over-aging
(slower dynamics) at long timescales. In addition, we investigate the
microscopic dynamics behind these relaxation phenomena through
measurements of the two-time particle correlation functions and the
van Hove distribution functions.

\section{Methodology}

We perform molecular dynamics (MD) simulations on a ``bead-spring''
polymer model that has been studied extensively for its glass-forming
properties \cite{Kremer_Grest,Binder_PRE57}. Beads interact via a
non-specific van der Waals potential (Lennard-Jones), and bonds are
modeled as a stiff spring (FENE) that prevents chain crossing. The
reference length-scale is $a$, the diameter of the bead; the energy
scale, $u_0$, is determined by the strength of the van der Waals
potential; and the time scale is $\tau_{LJ}=\sqrt{ma^2/u_0}$, where
$m$ is the mass of a bead.

\begin{figure}
	\centering \includegraphics{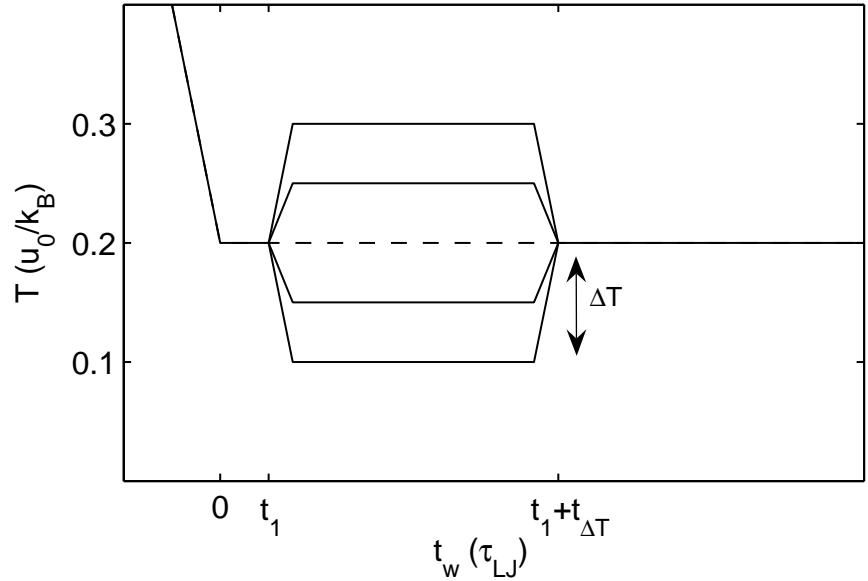}
	\caption{Simulation protocol for a temperature square
	step. $\Delta T =$ -0.1, -0.05, 0, 0.05, and 0.1$u_0/k_B$.}
	\label{fig:experiment}
\end{figure}

Our simulation method is similar to the one discussed in
ref.~\cite{Warren_PRE07}. We simulate 85500 beads in a cubic box with
periodic boundary conditions. All polymer chains have 100 beads
each. This length is not much greater than the entanglement length;
however as the dynamics in glasses are very slow, reptation effects
are minimal over the timescale of our simulations. The thermal
procedure is detailed in Fig.~\ref{fig:experiment}, and is very
similar to the protocol used in ref.~\cite{Lequeux_JSM} except for the
obvious difference in timescales due to the limitations of molecular
dynamics simulations. The glass is formed by a rapid quench at
constant volume from an equilibrated melt at $T=1.2u_0/k_B$ to a
glassy temperature of $T=0.2u_0/k_B$ ($T_g\approx0.35u_0/k_B$ for this
model \cite{Rottler_PRE64}). It is then aged at zero pressure and
$T=0.2u_0/k_B$ for $t_1=150\tau_{LJ}$, when the temperature is ramped
to $T=(0.2+\Delta T)u_0/k_B$ over a period of $75\tau_{LJ}$, held
there for $t_{\Delta T}=750\tau_{LJ}$, and then returned to
$T=0.2u_0/k_B$ at the same rate. We monitor the dynamics of the system
at wait times $t_w$ since the initial temperature quench.

\section{Results}
 \subsection{Dynamics}
\begin{figure}
	\centering \includegraphics{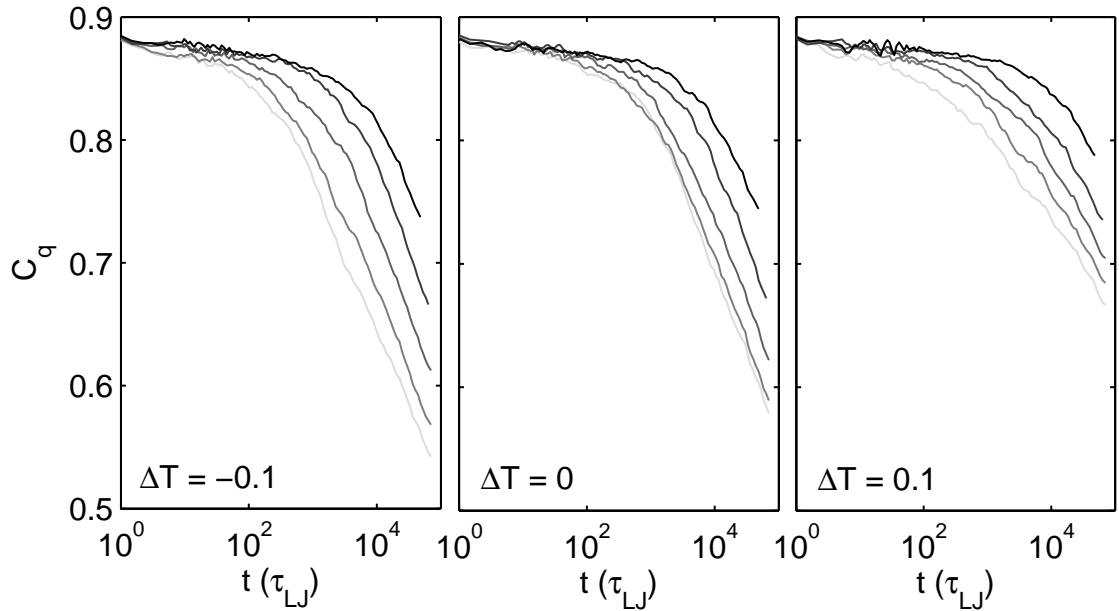}
	\caption{Incoherent scattering factor, $C_q(t,t_w)$, versus
	$t$ for various wait times $t_w$, and $q=6.3a^{-1}$. The
	curves from left to right in each panel have wait times of
	1125, 1800, 3300, 8550, and 23550$\tau_{LJ}$. }
	\label{fig:Cq_all}
\end{figure}

Two-time correlation functions are often used to measure the
structural relaxations in aging glassy systems. In this study, we
measure the incoherent scattering function,
\begin{equation}
C_q(t,t_w)=\frac{1}{N}\sum^{N}_{j=1}\exp(i\vec{q} \Delta \vec{r}_j(t,t_w))
\label{eqn:Cq}
\end{equation}
($\vec{q}$ is the wave vector, and $\Delta
\vec{r}_j(t,t_w)=\vec{r}_j(t_w+t)-\vec{r}_j(t_w)$ is the displacement
vector of the $j^{th}$ atom) as well as the mean squared displacement,
\begin{equation}
\langle{\Delta \vec r}(t,t_w)^2\rangle=\frac{1}{N}\sum^{N}_{j=1} \Delta \vec{r}_j(t,t_w)^2
\label{eqn:msd}
\end{equation}
as functions of the time $t$ after the wait time $t_w$ since the glass
was quenched. The two-time correlation functions have two regimes:
\begin{equation}
	C(t,t_w) = C_{fast}(t)+C_{slow}(t,t_w)
\end{equation}
$C_{fast}(t)$ describes the unconstrained motion of particles over
their mean-free path; $C_{slow}(t,t_w)$ describes the dynamics at much
longer timescales, where particle motions arise due to the collective
relaxations of groups of particles. Only the slow part of the
correlation function experiences aging, therefore discussion of the
aging dynamics of $C(t,t_w)$ refer to $C_{slow}(t,t_w)$.

In ref.~\cite{Warren_PRE07}, we showed that if $\Delta T = 0$, both
the incoherent scattering factor and the mean squared displacement
exhibit superposition with wait time: a simple rescaling of the time
variable by a shift factor $a$ causes complete collapse of the curves
at different wait times, i.~e.
\begin{equation}
C_q(t,t_w)=C_q(at,t_w')
\label{eqn:superposition}
\end{equation}
and
\begin{equation}
\langle \Delta \vec{r}(t,t_w)^2\rangle=\langle \Delta \vec{r}(at,t_w')^2\rangle.
\label{eqn:superposition2}
\end{equation}
The shift factor varies with the wait time in a simple power law,
\begin{equation}
a{\sim}t_w^{-\mu}
\label{eqn:aging}
\end{equation}
and agrees exactly with the shifts obtained from superimposing the creep
compliance curves. Such a power law in the shift factors is
characteristic of the aging process under a simple quench; however,
ref.~\cite{Lequeux_JSM} found that the power law does not hold for a
temperature square step protocol. Figure \ref{fig:Cq_all} shows
$C_q(t,t_w)$ for three temperature jumps, $\Delta T =$ -0.1, 0 and
0.1$u_0/k_B$. All curves exhibit a flat region at short times where
atoms are relatively immobile due to the caging effect of their
neighbours, followed by a rapid roll-off at longer times where atoms
begin to escape from the local cages. For the negative temperature
step, the $C_q(t,t_w)$ curves have the same shape as the reference
curves for $\Delta T = 0$. Superposition with wait time continues to
apply and the curves are simply shifted forward in time with increasing
$t_w$, although the shifts are clearly different from the reference
case. A positive temperature jump, however, causes a notable
modification of the shape of the correlation function. The curves roll
off more slowly at short wait times than long wait times, and
superposition with wait time is no longer possible.

\begin{figure}
	\centering
		\includegraphics{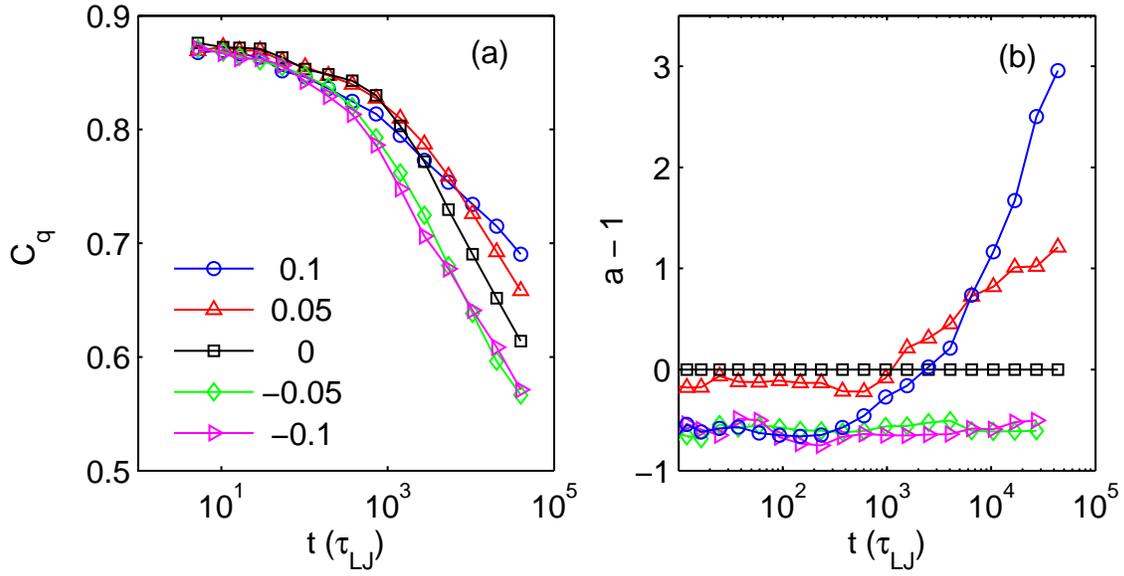}
	\caption{(a) Incoherent scattering function for various
	temperature steps (indicated in the legend in units of
	$u_0/k_B$). (b) Shift factors with respect to the $\Delta T =
	0$ case (Eq.~(\ref{eqn:shift_factor})) for the curves in (a).}
	\label{fig:Cq_teff}
\end{figure}

The changes in the relaxation time spectrum due to a temperature step
can clearly be seen in Fig.~\ref{fig:Cq_teff}(a), where $C_q(t,t_w)$
is plotted for each of the temperature steps at a wait time of
$1125\tau_{LJ}$. Compared to the case with $\Delta T = 0$, a step down
in temperature causes a constant shift in the correlation function
toward shorter times. The glass is said to be ``rejuvenated''. A
positive temperature step, however, seems to cause faster relaxations
(rejuvenation) at short times, and slower relaxations (over-aging) at
long times. For each of these curves, we define a shift factor with
respect to the $\Delta T = 0$ sample that depends on time,

\begin{equation}
	C_q(t,t_w,T_0) = C_q\left(a\left( t \right)t,t_w,T_0+\Delta T\right).
	\label{eqn:shift_factor}
\end{equation}
This quantity is exactly analogous to the effective time,
$t_{eff}(t)$, that was used to analyze the changes in the creep
compliance curves after a temperature square step in
ref.~\cite{Lequeux_JSM}:
\begin{equation}
	\frac{t_{eff}}{t_w}=a(t)-1.
\end{equation}
$a(t)-1$ is plotted in Fig.~\ref{fig:Cq_teff}(b) for the correlation
data in Fig.~\ref{fig:Cq_teff}(a). These results for the incoherent
scattering factor are in excellent qualitative agreement with the data
obtained from the experimental creep compliance \cite{Lequeux_JSM}. A
step down in temperature causes a relatively constant, negative shift
in the relaxation times, whereas a step up in temperature shows an
effective time that is negative for short times, but eventually
transitions to positive values for longer times.

\subsection{Negative temperature step}

\begin{figure}
	\centering
		\includegraphics{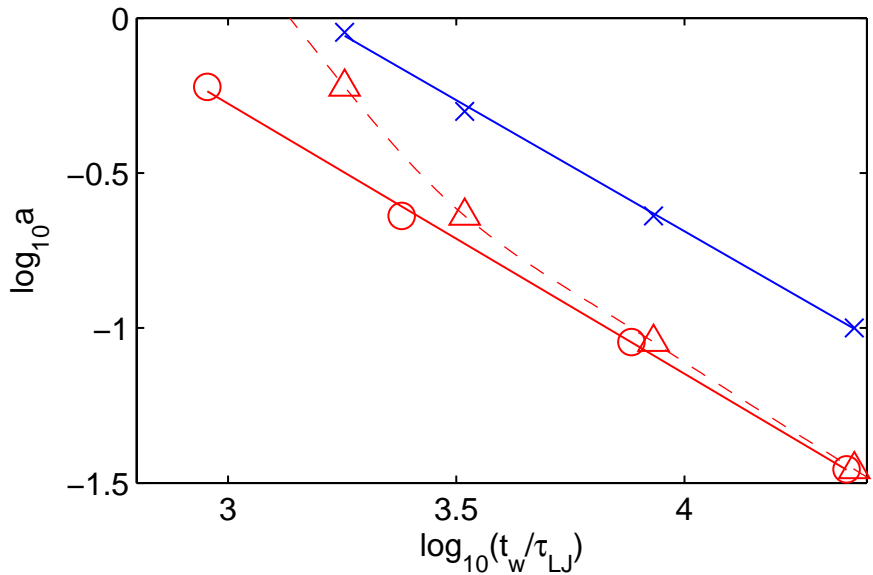}
	\caption{Shift factors (Eq.~\ref{eqn:superposition}) versus
	wait time for the $\Delta T = 0$ sample ($\times$), the
	$\Delta T = -0.1u_0/k_B$ ($\triangle$) sample, and the $\Delta
	T =-0.1u_0/k_B$ sample with $t_w'=t_w-t_{\Delta T}$
	($\ocircle$). The solid lines are power law fits to the data,
	and the dashed line is a guide to the eye.}
	\label{fig:shifts_downstep}
\end{figure}

As superposition of the incoherent scattering factor with wait time
continues to hold after a step down in temperature, we can monitor the
aging dynamics through the standard procedure of rescaling the time
variable to make a master curve of the correlation functions. Figure
\ref{fig:shifts_downstep} shows the shift factors versus wait time for
the data from Fig.~\ref{fig:Cq_all} for $\Delta T = 0$ and $\Delta T =
-0.1u_0/k_B$. The sample with no temperature step ages with the
characteristic power law in $t_w$, however, the sample that
experienced a negative temperature step clearly does not. However, if
we correct the wait time using the assumption that there is no aging
at the lower temperature, $t_w' = t_w-t_{\Delta T}$, then the power
law is restored. The aging exponents for these samples agree within
the uncertainty of the fit at $\mu=0.86\pm0.09$. A downward
temperature step of $\Delta T = -0.1u_0/k_B$ seems to induce a trivial
rejuvenation brought about by effectively ``freezing'' the dynamics at
the lower temperature.

\subsection{Positive temperature step}

\begin{figure}
	\centering
		\includegraphics{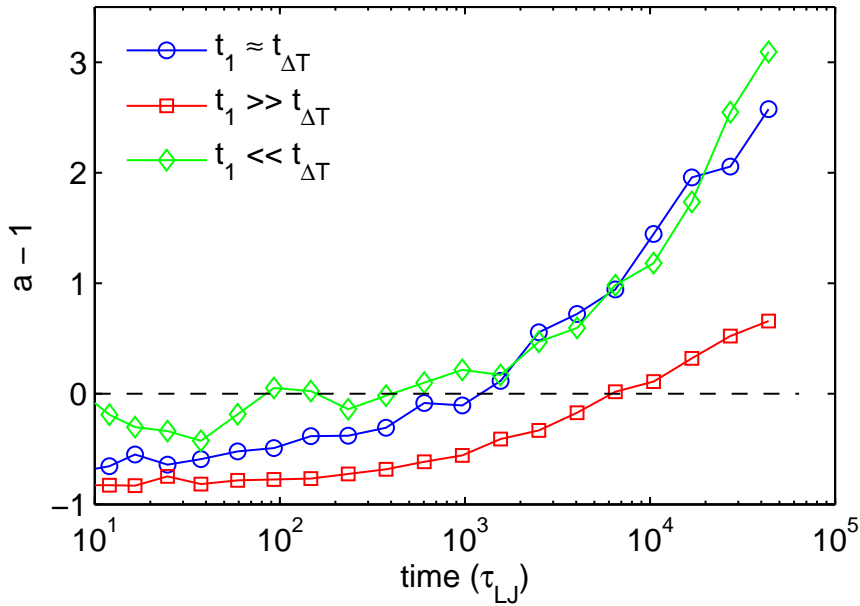}
	\caption{Shift factors versus time with respect to the
	unperturbed sample at $t_w = 1800\tau_{LJ}$ for samples with
	temperature jumps defined by $\Delta T = 0.1u_0/k_B$, and
	$(t_1,t_{\Delta T})$ equal to 825 and 750$\tau_{LJ}$
	($\ocircle$), 1313 and 263$\tau_{LJ}$ ($\Box$), and 263 and
	1313$\tau_{LJ}$ ($\Diamond$).}
	\label{fig:teff_vs_t1}
\end{figure}

The effect of an increase in temperature cannot be described as either
simple rejuvenation or over-aging, but instead the dynamics are
accelerated at short times and slowed at long times. To understand the
origin of the rejuvenation to over-aging transition, the relevant
timescales in the protocol, $t_1$ and $t_{\Delta T}$, are varied for a
constant step up in temperature of $0.1u_0/k_B$. We investigate three
cases: (1) $t_1 \approx t_{\Delta T}$, (2) $t_1 \gg t_{\Delta T}$ and
(3) $t_1 \ll t_{\Delta T}$. The effective time with respect to the
sample with no temperature step is shown for these simulation
parameters at $t_w=1800\tau_{LJ}$ in Fig.~\ref{fig:teff_vs_t1}. To
simplify the analysis, $t_1$ and $t_{\Delta T}$ were chosen such that
in each of the three cases, the dynamics are measured at the same time
since the end of the step. In cases (1) and (2), we clearly see
rejuvenation of the short time region of the relaxation-time spectrum
and a transition to over-aging at longer times; however, the
transition from rejuvenation to over-aging occurs later for case (2)
(shorter $t_{\Delta T}$). In case (3), there is no clear rejuvenation
region at all, and at long times, the relaxation spectrum is almost
identical to case (1).

These results may be understood based on the transient dynamics at the
higher temperature. The aging dynamics in glassy systems consist of
spatially and temporally heterogeneous relaxation events, whereby
collective motions in a group of particles lead to small
rearrangements of the cage \cite{Weeks_JCM15,Cipelletti_JPCM17}. These
relaxations are often called dynamical heterogeneities
\cite{Kob_PRL79,Vollmayr_PRE72}. After a first relaxation event (cage
escape), it has been shown that the timescale for subsequent
relaxations is shorter, and a typical group of atoms will experience
many such events before finding a stable configuration
\cite{Kob_PRL99}. Because the aging dynamics are thermally activated,
immediately after the temperature of the glass is raised, the total
rate of relaxation events rapidly increases and persists for a time
which is related to the ``life-time'' of the mobile regions. Once
these regions finally stabilize, the resulting structure has a lower
energy than before, or an over-population of long relaxation
times. Overall, as also pointed out in ref.~\cite{Lequeux_JSM}, the
greater rate of transitions at this temperature leads to accelerated
aging with respect to a glass at the lower temperature.  The results
of our simulations after a temperature square step can then be
understood as a competition between the initial rejuvenation due to
the dynamical heterogeneities, and the over-aging that results when
once they have stabilized. Therefore, a short temperature square step
causes mostly rejuvenation, and long temperature steps show primarily
over-aging.  This picture is generally consistent with that of trap models \cite{Bouchaud_PRL2003} that generate glassy dynamics through a wide distribution of relaxation times. However, ref.~\cite{Lequeux_JSM} found that such models do not explain all aspects of the thermal cycling experiment since they do not address the spatial arrangement of the relaxation processes.

It is curious that the long time part of the relaxation
spectrum is the same for case (1) and case (3). This may indicate the
end of the rapid transient effects, and the beginning of more steady
aging in the glass at the higher temperature. It would be useful to
expand the study to even longer $t_{\Delta T}$ to investigate this
finding.

\begin{figure}
	\centering
		\includegraphics{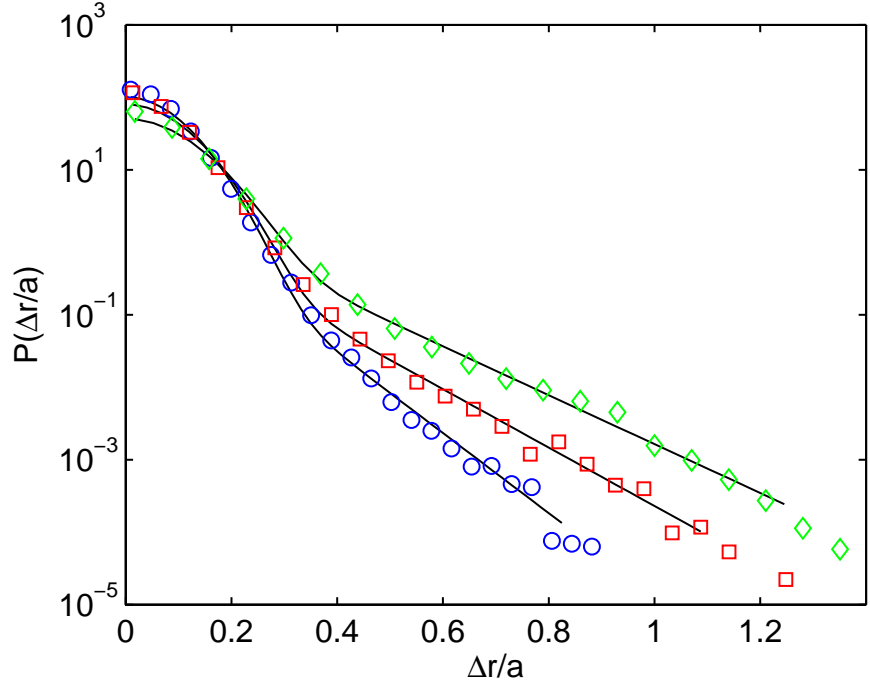}
	\caption{The van Hove function for a glass aged at
	$T=0.2u_0/k_B$ (no temperature jump) for
	$t_w=1125\tau_{LJ}$. The curves are for times of $15\tau_{LJ}$
	($\ocircle$), $150\tau_{LJ}$ ($\Box$) and $1575\tau_{LJ}$
	($\Diamond$). The black lines are fits to the curves using
	Eq.~(\ref{eqn:dist_fit}).}
	\label{fig:dist_nojump_vs_t}
\end{figure}

\subsection{Particle displacement distributions}

In addition to the correlation functions, which provide an average
picture of the particle dynamics with $t$ and $t_w$, molecular
dynamics simulations allow us to obtain the full distribution of
displacements $P(\Delta r,t,t_w)$, where $\Delta r = |\Delta
\vec{r}|$. This is also called the van Hove function
\cite{Kob_PRE51}. The van Hove function was measured for three
temperature jumps ($\Delta T=$ -0.1, 0 and 0.1$u_0/k_B$), and
representative curves are shown in Fig.~\ref{fig:dist_nojump_vs_t} for
$\Delta T = 0$. The distribution appears to be the
combination of a narrow, caged particle distribution, and a wider
distribution of ``mobile'' particles that have experienced a cage
rearrangement. The distribution can be described by the sum of a
Gaussian for the caged particles, and an exponential tail for the
mobile particles:
\begin{figure}
	\centering
		\includegraphics{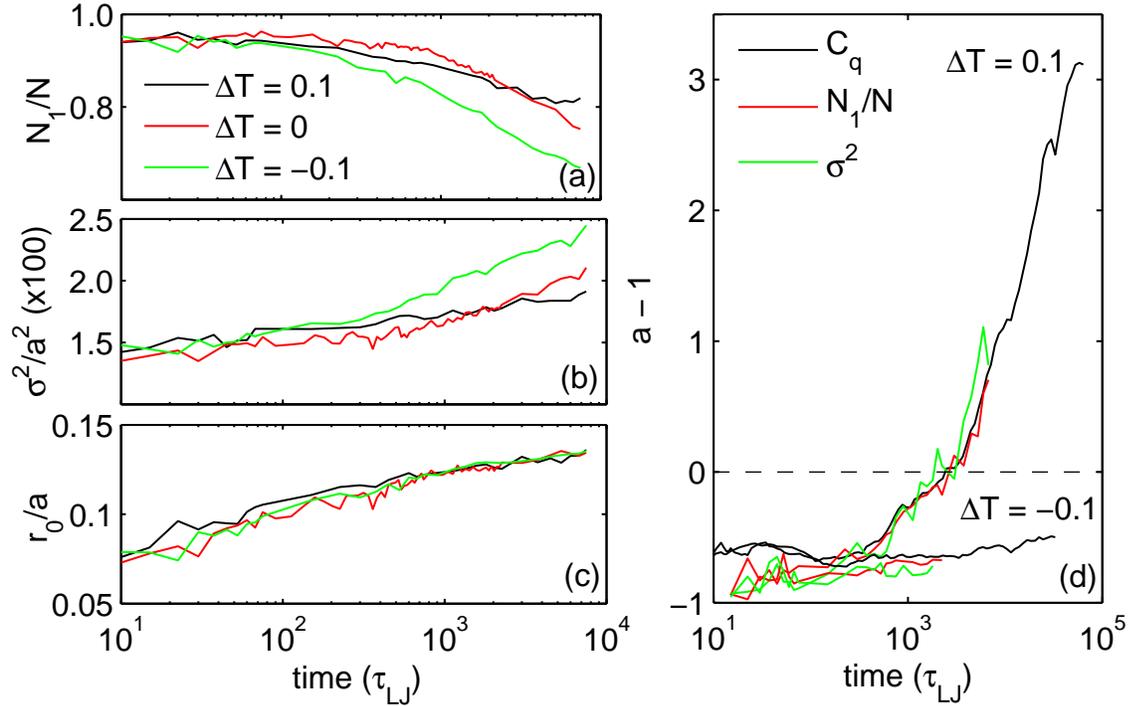}
	\caption{(a)-(c): Fit parameters to the van Hove distributions
	(Eq.~(\ref{eqn:dist_fit})) for three different temperature
	jumps at $t_w=1125\tau_{LJ}$. (d): Shift factors for $N_1/N$
	and $\sigma^2$ together with those from $C_q(t,t_w)$ obtained
	from Fig.~\ref{fig:Cq_teff}.}
	\label{fig:teff_fit_params}
\end{figure}
\begin{equation}
	P(\Delta r) = N_1 e^{-\Delta r^2/\sigma^2} + N_2 e^{-\Delta r/r_0}.
	\label{eqn:dist_fit}
\end{equation}
If the distribution is normalized, there are three fit parameters:
$N_1/N$, the ratio of caged particles $N_1$ to total particles $N=N_1+N_2$; $\sigma^2$, the width
of the cage peak; and $r_0$, the characteristic length-scale of the
exponential tail. These are shown in
Fig.~\ref{fig:teff_fit_params}(a)-(c). $N_1/N$ is relatively flat at
short times, followed by a decay in the number of trapped particles
that signals the onset of cage escape. The width of the cage peak,
$\sigma^2$, is constant on the timescale where the particles are
predominantly caged, and increases sub-diffusively in the cage escape
regime possibly due to weak coupling between adjacent cages (local
relaxations may somewhat affect cages nearby
\cite{Heuer_PRE72}). Alternatively, the length scale of the mobile
distribution, $r_0$, increases steadily with time even at the shortest
timescales. The parameters $N_1/N$ and $\sigma^2$ show large
differences for samples undergoing different temperature jumps,
however, $r_0$ does not. A previous study showed that the width of the
mobile peak also did not depend on the wait time
\cite{Warren_PRE07}. It seems that the shape of the mobile
distribution does not exhibit memory.

\begin{figure}
	\centering
		\includegraphics{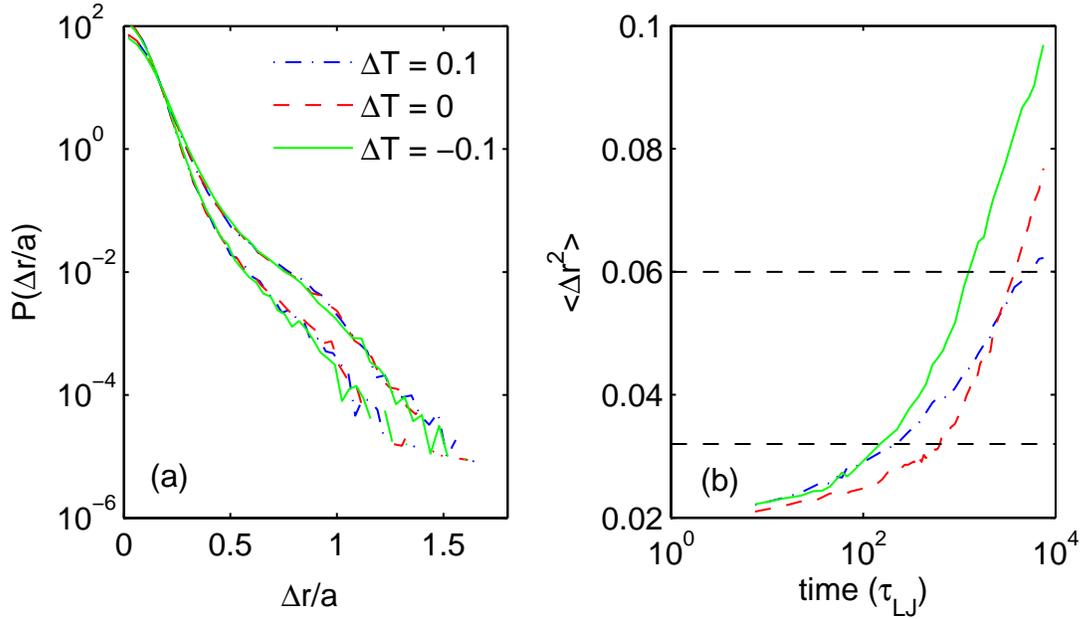}
	\caption{(a) The van Hove distributions for three temperature
	steps (see legend) at times where their mean squared
	displacement are equal. (b) The mean-squared displacement
	versus time for the temperature steps in (a). The intercepts
	of the dashed lines and the mean-squared displacement curves
	indicate the measurement times of the distributions in (a).}
	\label{fig:dist_msd_t1}
\end{figure}

The effects of the temperature step on the van Hove function can be
understood by defining a time-dependent shift factor for $\sigma^2$
and $N_1/N$, in analogy to Eq.~(\ref{eqn:shift_factor}) for
$C_q(t,t_w)$. Surprisingly, we see in
Fig.~\ref{fig:teff_fit_params}(d) that the shift factors for both fit
parameters and for $C_q(t,t_w)$ are identical. This suggests that the
van Hove functions after different temperature steps can be
superimposed at times where their two-time correlation functions are
equal,
\begin{equation}
	P(\Delta r,t,t_w) = P(\Delta r, \langle \Delta r(t,t_w)^2 \rangle).
	\label{eqn:P_superpos}
\end{equation}
Indeed, this is what we observe in Fig.~\ref{fig:dist_msd_t1}(a). The
mean-squared displacement for this system is shown
Fig.~\ref{fig:dist_msd_t1}(b); the distributions are compared at times
indicated by the intercept of the $\langle \Delta r^2 \rangle$ curves
with the dashed horizontal lines. Note that this is not a simple
rescaling of the length parameter $\Delta r/\sqrt{\langle \Delta r^2
\rangle}$, as there are two distinct length-scales in the system,
$\sigma$ and $r_0$. Superposition of the van Hove function has also
been found for variations in wait time $t_w$ after a simple quench
from the melt \cite{Castillo_NaturePhys3,Warren_PRE07}. For this
thermal protocol, the entire spectrum of relaxation times scales as
$t/t_w^\mu$, in which case a simple Landau-theory approximation
predicts full scaling of the probability distributions
\cite{Castillo_PRL88}. Full scaling of the relaxation times clearly
does not hold after a temperature step, therefore it can not be a
necessary condition for superposition of the van Hove
distributions. It seems that the relationship described in
Eq.~(\ref{eqn:P_superpos}) is quite general.

Superposition of the van Hove functions may be a consequence of the
fact that, while the relaxation times are greatly modified by wait
time and temperature, the spatial scale of the relaxations is
not. This microscopic length scale is determined by the density, which
is known to be much less sensitive to these parameters. Although
correlation lengths of rearranging clusters can be quite large in
glasses, these lengths also do not appear to grow with aging
\cite{Weeks_JCM15}. If the typical size of relaxation events is
approximately constant, then we can surmise that the only relevant
quantity in the van Hove function is the number of relaxation events.
This information is exactly expressed by the effective time that we
obtained from the average correlation functions. Future work will
investigate the particle trajectories in more detail to validate the
spatial scale of the cage rearrangements and its relationship to
the microscopic length scale of the system.

\section{Conclusion}

Molecular dynamics simulations of physical aging in a model polymer
glass were performed using a thermal protocol modeled after recent
experiments \cite{Lequeux_JSM}. A temperature square step was applied
to the glass after an initial quench from the melt, and the dynamics
were monitored through the two-time correlation functions. Results
show excellent agreement with experiment. A negative temperature step
causes uniform rejuvenation due to reduced aging at low
temperatures. A positive temperature step yields a completely
different relaxation spectrum: at short times the dynamics are
accelerated (rejuvenation), and at long times they are slowed
(over-aging). By modifying the length of the step up in temperature,
we determined that the transition from rejuvenation to over-aging is
controlled by the length of the square step: short steps cause
primarily rejuvenation, while long steps show marked over-aging. We
also investigated the distribution of displacements,
or the van Hove functions. Even though the spectrum of relaxation
times was greatly modified by the temperature step, the van Hove
functions showed perfect superposition at times where the mean squared
displacements were equal.

\section{Acknowledgements}
We thank the National Sciences and Engineering Council of Canada
(NSERC) and the Canadian Foundation for Innovation (CFI) for financial
support. Computing resources were provided by WestGrid.

\end{document}